%% file: main.tex
\documentclass[runningheads,table]{llncs}

\usepackage{eccv}

\usepackage{eccvabbrv}
\usepackage[export]{adjustbox}

\usepackage{graphicx}
\usepackage{booktabs}

\input{preamble}
\usepackage[accsupp]{axessibility}  

\usepackage[pagebackref,breaklinks,colorlinks]{hyperref}

\usepackage{orcidlink}

\begin{document}

\title{Hypernetworks for Generalizable BRDF Representation} 

\titlerunning{Hypernetworks for Generalizable BRDF Representation}

\author{Fazilet Gokbudak\inst{1} \and
Alejandro Sztrajman\inst{1} \and
Chenliang Zhou\inst{1}
Fangcheng Zhong\inst{1} \and
Rafal Mantiuk\inst{1}
Cengiz Oztireli\inst{1}
}
\authorrunning{F.~Gokbudak et al.}

\institute{University of Cambridge, Cambridge, UK \and
Department of Computer Science and Technology,
The Computer Laboratory, William Gates Building
15 JJ Thomson Avenue, Cambridge CB3 0FD
\url{https://www.cst.cam.ac.uk/}\\
\email{fg405@cam.ac.uk}}


\maketitle

\begin{abstract}
In this paper, we introduce a technique to estimate measured BRDFs from a sparse set of samples. Our approach offers accurate BRDF reconstructions that are generalizable to new materials. This opens the door to BDRF reconstructions from a variety of data sources. The success of our approach relies on the ability of hypernetworks to generate a robust representation of BRDFs and a set encoder that allows us to feed inputs of different sizes to the architecture. The set encoder and the hypernetwork also enable the compression of densely sampled BRDFs. We evaluate our technique both qualitatively and quantitatively on the well-known MERL dataset of 100 isotropic materials. Our approach accurately 1) estimates the BRDFs of unseen materials even for an extremely sparse sampling, 2) compresses the measured BRDFs into very small embeddings, e.g., 7D.

\end{abstract}

\input{sec/1_introduction}
\input{sec/2_relatedWork}
\input{sec/3_method}
\input{sec/4_experiments}
\input{sec/5_conclusion}

%
%
\bibliographystyle{splncs04}
\bibliography{main}

\end{document}


\title{Hypernetworks for Generalizable BRDF Representation} 

\titlerunning{Abbreviated paper title}

\author{First Author\inst{1}\orcidlink{0000-1111-2222-3333} \and
Second Author\inst{2,3}\orcidlink{1111-2222-3333-4444} \and
Third Author\inst{3}\orcidlink{2222--3333-4444-5555}}

\authorrunning{F.~Author et al.}

\institute{Princeton University, Princeton NJ 08544, USA \and
Springer Heidelberg, Tiergartenstr.~17, 69121 Heidelberg, Germany
\email{lncs@springer.com}\\
\url{http://www.springer.com/gp/computer-science/lncs} \and
ABC Institute, Rupert-Karls-University Heidelberg, Heidelberg, Germany\\
\email{\{abc,lncs\}@uni-heidelberg.de}}

\maketitle

\input{sec/X_suppl}

%
%
\bibliographystyle{splncs04}
\bibliography{main}

%% file: preamble.tex
\usepackage[dvipsnames]{xcolor}

\usepackage{array}
\usepackage{booktabs}

\makeatletter
\newcommand{\thickhline}{%
    \noalign {\ifnum 0=`}\fi \hrule height 1pt
    \futurelet \reserved@a \@xhline
}
\newcolumntype{"}{@{\hskip\tabcolsep\vrule width 1pt\hskip\tabcolsep}}
\makeatother


%% file: sec/1_introduction.tex
\section{Introduction}
\label{sec:intro}


In computer graphics and vision, achieving realistic renderings for intricate surface materials hinges on accurately describing the interaction of light with the surfaces. This is conventionally conveyed through the modeling and reconstruction of a 4D Bidirectional Reflectance Distribution Function (BRDF), which quantifies the relationship between the incident and outgoing light intensities for a specific material. In this work, we propose a novel generalizable BRDF representation model that can estimate the BRDFs of new materials from highly sparse and unstructured point-based samples \footnote{Point-based samples refer to the BRDF values acquired at specific points on a surface with known viewing and lighting directions.} and compress the densely sampled values into very small latent embeddings.



While there has been prior work that attempts to tackle similar tasks by estimating the free parameters of analytic BRDF functions or the principal components of BRDF functions from images or reflectance measurements, the oversimplified models of the complex BRDF function lead to inaccuracies when predicting real materials, thereby diminishing the realism in renderings~\cite{ngan2005}. Moreover, the process of fitting through nonlinear optimization is inherently unstable, computationally expensive, and prone to local minima, hindering the accurate reconstruction of material appearance~\cite{dupuy2015, guarnera2016}. Measured BRDFs of real-world materials can also be fraught with errors due to equipment limitations, adding to the complexity of the fitting process~\cite{nielsen2015optimal}. 
\begin{figure}
  \centering
   \includegraphics[width=\linewidth]{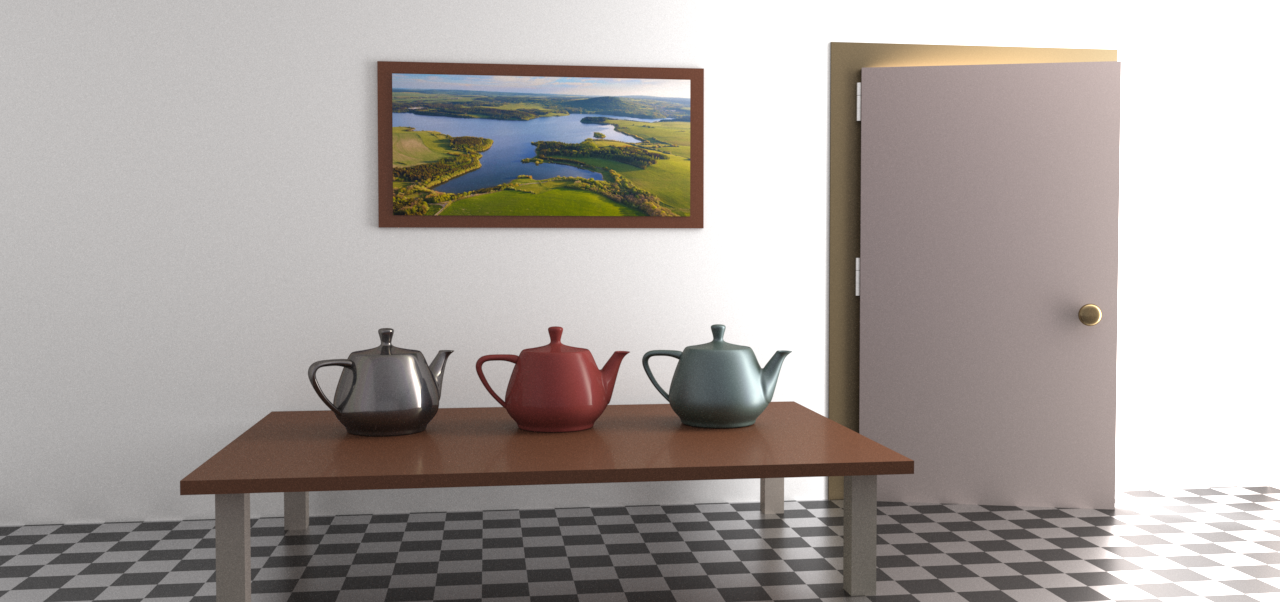}
   \caption{A room scene rendered with our reconstructed materials, including sparse reconstruction (table top and legs, door, door and picture frames, hinge), compression (two teapots on the left, door handle) and BRDF interpolation (right-most teapot). Scene courtesy of Benedikt Bitterli.}
   \label{fig:teaser}
\end{figure}

Recent advances in deep learning have enabled the accurate representation of complex continuous signals (\textit{e.g.} images, surfaces, volumes, materials, \textit{etc.}) using a Neural Field~\cite{sitzmann2020siren, ffn, cnf2023}, \textit{i.e.} a neural network mapping coordinate inputs to sampled values, without compromising model compactness. Consequently, neural fields have gained substantial popularity for representing continuous BRDFs in recent works~\cite{sztrajman2021neural, cnf2023}. However, reconstructing a neural field representation for BRDFs typically requires training the network with a regression loss function to overfit to a single material, which demands dense sampling and extensive computational resources, while being unable to generalize to new materials.

More recently, Generalizable Neural Fields~\cite{rebain2022attention} (GNFs) have emerged as a promising solution to the aforementioned challenges. Rather than overfitting to individual signals, GNFs are designed to learn a generalized mapping, either deterministic or probabilistic, between sampled signals and their full neural field representations in a fashion similar to supervised learning.
The key strategy of GNFs involves conditioning the neural fields with a latent embedding of the signal samples. Popular conditioning mechanisms include concatenated latent vectors~\cite{park2019deepsdf}, hypernetworks~\cite{ha2017hypernetworks}, and attention-based set latent representations~\cite{jiang2021cotr}. Once the conditional neural field is properly learned, reconstruction of the full signal can be achieved at inference time, and even with highly sparse and unstructured samples.


Inspired by GNFs, we propose a novel framework for generalizable neural BRDF representation for both the estimation of unseen materials from sparse and unstructured samples and the compression of measured BRDFs into low-dimensional latent space. In this framework, we employ a multi-layer perceptron (MLP) model as the neural field backbone for BRDF representation, a hyper-network for conditioning, and a set encoder that allows for mapping an arbitrary number of reflectance measurements from arbitrary directions to a compact latent embedding for conditional BRDF generation. 
The built-in nonlinear interpolation capability of the hyper-network also offers robust and adaptable material editing and blending across various sample sizes.

Unlike previous work, our BRDF reconstruction is highly efficient without the need for extensive training to overfit individual materials, while also maintaining state-of-the-art performance in reconstruction accuracy for sample size below 4000; see Figure~\ref{fig:imp_comp_upt}.

In summary, we offer a novel solution for the challenging task of generalizable BRDF modelling with the following key contributions:
\begin{itemize}
    \item{\textbf{Generalizability and Adaptability:} Our methodology ensures robustness and adaptability across varying sample sizes and appearances, making it highly effective for estimating the BRDFs of unseen materials from highly sparse and unstructured sampling. Its adaptability also extends to highly compact representation of BRDFs, overall outperforming the prior state-of-the-art compression method; see Table \ref{table: oursvsnps}.}
    

    \item{\textbf{Realism and Accuracy:}
    Our extensive evaluation demonstrates the superior performance of our approach in reconstructing the BRDFs of 20 test materials from a limited number of samples, ranging from 40 to 4000, outperforming prior methods in terms of appearance modeling and color preservation (by at least 2dB in peak-signal-to-noise ratio and 1 in Delta E).
    }
\end{itemize}

%% file: sec/2_relatedWork.tex
\section{Related Work}
\label{sec:relatedwork}



\subsection{Analytic BRDF Models}
Analytic models are the most common representation for BRDFs. Classic BRDF models include Phong~\cite{blinn77}, Cook-Torrance~\cite{cooktorrance1982}, Ward~\cite{ward1992} and GGX~\cite{walter2007microfacet}. Following models have increased their reconstruction capabilities by combining analytic formulations with data-driven representations of some or all of their components~\cite{dupuy2015, ashikhmin2007, bagher2016}. Notably, the ABC model~\cite{low2012} has been shown to provide an accurate reconstruction of measured materials, while only requiring the fitting of a handful of tunable parameters. These models are usually fast at evaluation, easily editable, and present a low memory footprint. However, they usually rely on oversimplifications of the reflectance distribution shapes, and thus have a limited capacity for the reconstruction of complex real-world materials~\cite{ngan2005, guarnera2016}. Therefore, recent works have started exploring neural representations to overcome these limitations.






\subsection{Regression-based BRDF Estimation}
For a simpler representation of measured BRDFs, BRDF decomposition methods, such as PCA decomposition 
\cite{matusik2003data, nielsen2015optimal, serrano2018intuitive}, non-negative matrix factorization \cite{lawrence2004efficient, lawrence2006inverse}, Gaussian mixture \cite{sun2007interactive}, tensor decomposition \cite{bilgili2011general, tongbuasirilai2020compact} 
and non-parametric models~\cite{bagher2016non} have been proposed. The main limitation of PCA and factorization methods is  that they have a limited capacity to represent complex functions without overfitting to the training dataset. In contrast, our method can represent the complex BRDFs even with sparse samples while maintaining generalizability.




\paragraph{Deep learning for BRDF modeling.}

Multiple methods have been recently proposed for neural BRDF representation~\cite{rainer2019neural, hu2020deepbrdf, sztrajman2021neural, zheng2021compact, maximov2019deep, chen2021invertible, fan2021neural, cnf2023}. These methods usually offer a flexible representation, and thus are well fitted to encode the complex reflectance distributions of real-world measured materials. However, an accurate fitting of these methods usually requires lengthy optimizations and a very large number of sample measurements, typically from $8 \times 10^5$ to $1.5 \times 10^6$. \cite{maximov2019deep} learned materials with baked illumination via small fully-connected networks. NBRDF \cite{sztrajman2021neural} and CNF~\cite{cnf2023} leveraged neural fields to learn individual BRDF functions. Closer to our work, DeepBRDF~\cite{hu2020deepbrdf} and Neural Processes~\cite{zheng2021compact} introduce neural network architectures to learn a compressed latent space from a dataset of multiple materials. However, these methods only address the problem of compressing BRDF samples into a low dimensional space, hence overfitting to the dataset. Our method, on the other hand, also offers a generalizable approach for the reliable reconstruction of unseen materials from sparse and unstructured real-world reflectance measurements.

\subsection{Efficient BRDF Acquisition}
Realistic reflectance acquisition commonly requires a large amount of physical acquisition samples collected from different directions, making the process time-consuming and data intensive. To take fewer samples, hence reducing the BRDF capture time, \cite{nielsen2015optimal} propose optimizing for a sample pattern with a linear statistical analysis of a database of BRDFs. Along this line, \cite{liu2023learning} jointly learn the sample pattern and a non-linear BRDF model.

\paragraph{Spatially-varying BRDFs (SVBRDF):} For efficient SVBRDF capture, several methods based on multiplexing-based, also known as illumination-based, acquisition systems \cite{kang2018efficient, kang2019learning, ma2021free, ma2023opensvbrdf, tunwattanapong2013acquiring} have been proposed. A common approach has been to optimize the lighting patterns for efficient acquisition, followed by a BRDF fitting to an analytic model. Recent works have also leveraged deep learning architectures to learn a mapping from images to texture maps of analytic SVBRDF parameters~\cite{guo2021highlight, hui2017reflectance, deschaintre2018single, deschaintre2019flexible, martin2022materia, zhou2021adversarial,gao2019deep}. 

In contrast to those works, our focus is the reconstruction of spatially uniform BRDF that can accurately represent arbitrary complex materials. The works on spatially-varying BRDFs and efficient capture could be considered orthogonal to ours, and those methods could be potentially combined with ours. 





\subsection{Hypernetworks and GNFs}
The capacity of hypernetworks to dynamically output neural network weights, which allows models to adjust to input conditions, has drawn attention. Its promise in a variety of computer vision tasks, including dynamic network adaptation and generating neural implicit fields, has been demonstrated by recent works, such as HyperGAN \cite{ratzlaff2019hypergan} and Hyperdiffusion \cite{erkocc2023hyperdiffusion}.
The concept of a generalizable neural field has also been extensively applied to the reconstruction of neural radiance fields~\cite{wang2022attention, yang2023contranerf}, but not sufficiently studied in other domains.
These research efforts serve as our source of inspiration as we apply hypernetworks and GNFs to BRDF estimation, improving the model's adaptability to various material appearances.


%% file: sec/3_method.tex
\section{Methodology}

\begin{figure*}[t]
  \centering
   \includegraphics[width=\linewidth]{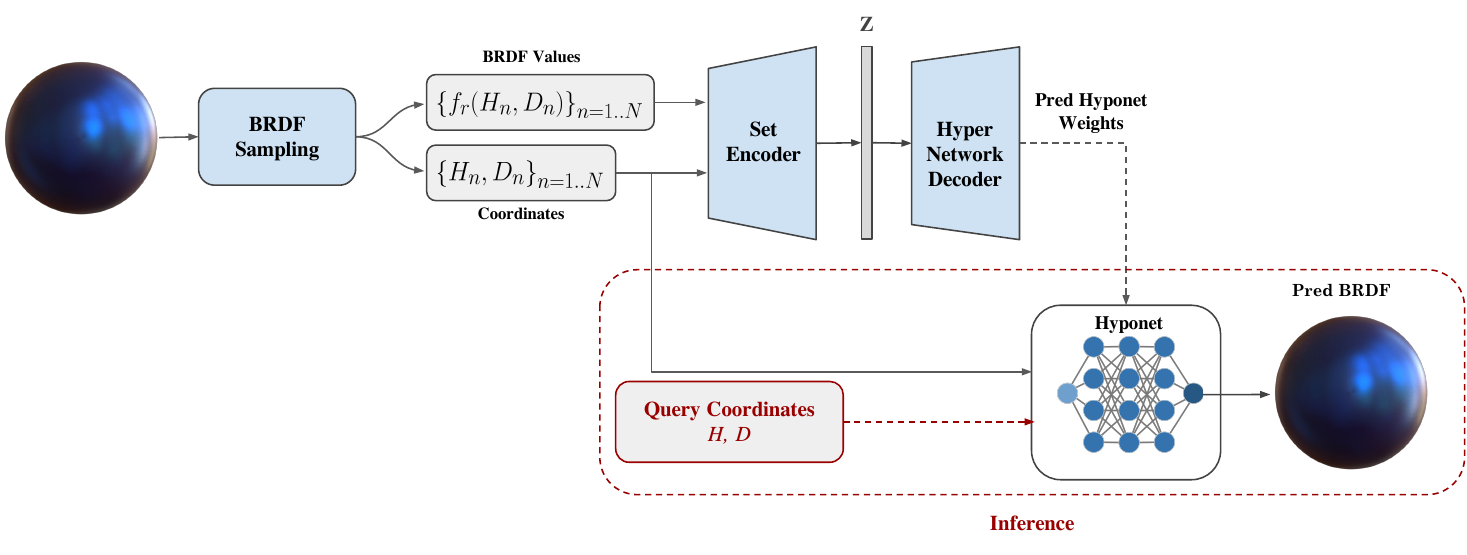}
   \caption{During training, the set encoder and hypernetwork decoder are trained on a set of materials to predict the weights of hyponet (MLP) so that it can reconstruct the training set. The BRDF data is provided as a set of BRDF coordinates, $H_n,D_n$, and the corresponding reflectance values $f_r(H_n,D_n)$. To reconstruct a new material from a small set of BRDF reflectance samples, the trained set encoder and hypernetwork decoder are used to predict the weights of hyponet for the unknown material. Once those weights are known, we can query BRDF at any coordinates and for any new materials, conditioned on the embedding of their sampled BRDF values.}
   \label{fig:mainfig} \end{figure*}



Based on our observation that recent work misses a generalized and adaptable BRDF representation, we propose a novel representation for measured BRDFs that learns compact embeddings of the BRDFs with a hypernetwork model.

\subsection{Pre-processing}\label{sec:pre-proc}

Measured BRDFs usually contain high dynamic range (HDR) data, including arbitrarily high values, especially for the specular components. 
Following~\cite{nielsen2015optimal}, we pre-process the BRDF data by applying a Log Relative Mapping of the form:
\begin{equation}
  \rho' = \ln{\left(\frac{\rho + \epsilon}{\rho_{ref} + \epsilon} +1\right)}
  \label{eq:preprocess}
\end{equation}
where $\rho$ refers to the BRDF values, $\epsilon = 0.002$ is a small constant value to avoid zero-division, and $\rho_{ref}$ is a reference BRDF value for relative mapping. As in \cite{nielsen2015optimal}, we choose the reference BRDF value to be the median value for each angle over the entire dataset.


We also observe that input parameterization has a strong impact on the reconstruction quality since it guides the neural network model to learn different aspects of the reflectance function. Therefore, similar to \cite{sztrajman2021neural}, we express the BRDFs as functions of the Cartesian vectors $H$ and $D$ in the Rusinkiewicz parameterization \cite{rusinkiewicz1998new},
\textit{i.e.}, $\rho=f_r(H, D)$, where $H, D \in S^2$ indirectly encode the information about the incident and outgoing light directions.
Importantly, in this parameterization the directions of specular reflection have a single fixed representation as $H=(0,0,1)$, which provides an easier pattern for the network to learn than the traditional $\omega_i, \omega_o$ encoding.


\subsection{Hypernetwork}
\label{sec:hypernet}

In Figure~\ref{fig:mainfig}, we show the diagram of our hypernetwork model~\cite{sitzmann2020siren} for BRDF representation, with three main components: 1) a set encoder that generates compact latent representations $Z$ of BRDFs based on an arbitrary number of directional samples, 2) a hypernetwork decoder that decodes the latent to estimate the parameters of a neural field,
and 3) a neural field controlled by the decoder that represents the BRDFs of the material, which we refer to as a hyponet, following the convention of prior work~\cite{sitzmann2020metasdf}.


\subsubsection{Set Encoder} 

The set encoder takes as input an arbitrary set of samples, $n=1..N$, taken from a BRDF measurement. Each measurement consists of directional coordinates ${H_n, D_n}$, given in the Rusinkiewicz parameterization~\cite{rusinkiewicz1998new}, and their corresponding BRDF values $f_r(H_n,D_n)$. The set encoder is composed of four fully-connected layers with two hidden layers of feature size 128 for each. The input is the concatenation of BRDF values $(N, 3)$ and coordinates $(N, 6)$. The activation function applied after each layer is ReLU. Each sample is encoded into a 40-dimensional embedding, and the sample set is reduced to a single embedding by applying a symmetric operation (averaging).
The use of a set encoder, which is commonly adopted in point set networks \cite{zaheer2017deepsets}, ensures permutation invariance and provides a high degree of flexibility in terms of the input, enabling the encoding of BRDFs with an arbitrary number of data-points, irregularly sampled, and in no pre-defined order.

\subsubsection{Hypernetwork Decoder and Hyponet} 
The hypernetwork decoder converts the embeddings from the set encoder into the weights of the hyponet that represents the BRDFs of a single material. The hypernetwork decoder is composed of 10 blocks of a fully-connected neural network with three layers. Each block outputs the corresponding weights and biases of hyponet. The hyponet consists of five fully-connected layers with input layer of size 6 for coordinates, three hidden layers of size 60 for each, output layer of size 3 for BRDF values. Our neural representation of materials, hyponet, follows the structure of Neural BRDF, as defined by~\cite{sztrajman2021neural}, but replaces the exponential activation in the last layer with a ReLU activation due to BRDF properties ($\rho \ge 0$). This network provides a continuous representation of a BRDF, and has been shown to provide state-of-the-art reconstructions of measured BRDFs, with performances competitive with the fastest analytic BRDF models.




\subsection{Training}
\label{sec:traindet}

We train the hypernetwork by optimizing the following loss, which consists of a reconstruction term $\mathcal{L}_\text{rec}$ and two regularization terms $\mathcal{L}_\text{weights}$ and $\mathcal{L}_\text{latent}$ for the hyponet weights $w$ and the latent embeddings $z$ ~\cite{ha2017hypernetworks}:
\begin{equation}
    \mathcal{L} = \mathcal{L}_\text{rec} +
              \lambda_1 \underbrace{\frac{1}{W} \sum_{j=1}^W w^2_j}_{\mathcal{L}_\text{weights}} +
              \lambda_2 \underbrace{\frac{1}{Z} \sum_{k=1}^Z z^2_k}_{\mathcal{L}_\text{latent}}
    \label{eq:loss}
\end{equation}

We define the reconstruction loss as the mean squared error between the cosine weighted predicted and ground-truth BRDF values:
\begin{equation}
    \mathcal{L}_{\text{rec}} = \sum_{n=1}^{N}\sum_{m=1}^{M}\frac{\left|\left|\rho^{\text{pred}}_{n, m} \cos{\theta_{n, m}} - \rho^{\text{true}}_{n, m} \cos{\theta_{n, m}}\right|\right|_{2}}{NM}
    \label{eq:Lrec}
\end{equation}
where $\rho^{\text{pred}}_{n, m}$ and $\rho^{\text{true}}_{n, m}$ indicate the predicted and ground truth BRDF values of the $n$-th sample of the $m$-th material, both processed as described in Eqn \ref{eq:preprocess}, and $\theta$ measures the angle between the incident ray and the surface normal. The cosine term weighs BRDFs based on the assumption of uniform incoming radiance and leads to more visually accurate results \cite{ngan2005experimental}.

We train our network for 80 epochs with $1\,458\,000$ samples per material. It takes around 15 minutes with an NVIDIA A100 80GB GPU support. 

\paragraph{Inference:}
When inferring the reflectance values, our model first estimates hyponet weights from sparse samples of a test material, which takes around 0.01 seconds without GPU. Later, we feed query coordinates to the hyponet to predict the BRDF values of the material. With the conversion of the predicted BRDF into a renderable format, this process takes around 9 seconds without GPU. The continuous representation of BRDFs with the hyponet offers a nonlinear built-in interpolation and, hence, accurately reconstruct unseen materials from even a few samples.


%% file: sec/4_experiments.tex
\section{Experiments}\label{sec:exp}

\subsection{Datasets and Baselines}

To show the effectiveness of our method, we use MERL \cite{Matusik2003jul} and RGL (51 isotropic materials) \cite{dupuy2018adaptive} datasets, which are the most commonly used BRDF datasets that include isotropic materials. The MERL dataset \cite{Matusik2003jul} consists of 100 real-world materials, covering a wide range of appearances. Each material includes reflectance measurements from a dense set of directions, parameterized as the spherical angles ($\theta$, $\phi$) of the $H$ and $D$ vectors from the Rusinkiewicz parameterization \cite{rusinkiewicz1998new}. Each color channel has a resolution of 90 × 90 × 180, leading to $1\,458\,000$ reflectance measurements. 




\subsection{Sparse BRDF Reconstruction}\label{sec:brdf_rec}
 For comparison with baselines, we train our hypernetwork model with 80 MERL materials that are randomly selected. We leave the remaining 20 materials for testing. To understand the reconstruction capacity of our model, we first train the model with all available samples ($1\,458\,000$). We observe that reducing the number of samples by around half ($640\,000$) results in a similar performance (Section 7 in supplementary). 

We also train our model with a mixed dataset of 80 MERL materials and 51 isotropic RGL materials. We later test the trained model on the same test dataset (20 MERL materials). We apply a separate log relative mapping to RGL materials by computing the median over the isotropic RGL dataset.
 
 We qualitatively compare our results against the ground truths through renderings of the materials. The renderings are obtained by a Mitsuba renderer with an environment map illumination. Our model can capture the diffuse colors of varying unseen appearances even for the materials with specular components.

The main advantage of our architecture is that it is flexible in the number of samples fed to the network. That is, we can reconstruct unseen materials with fewer samples than the sample number used during training. Thanks to its built-in nonlinear interpolation that comes from the hyponet, we obtain high quality reconstruction results with fewer samples. 

\begin{figure}[t]
  \centering
   \includegraphics[width=0.8\linewidth]{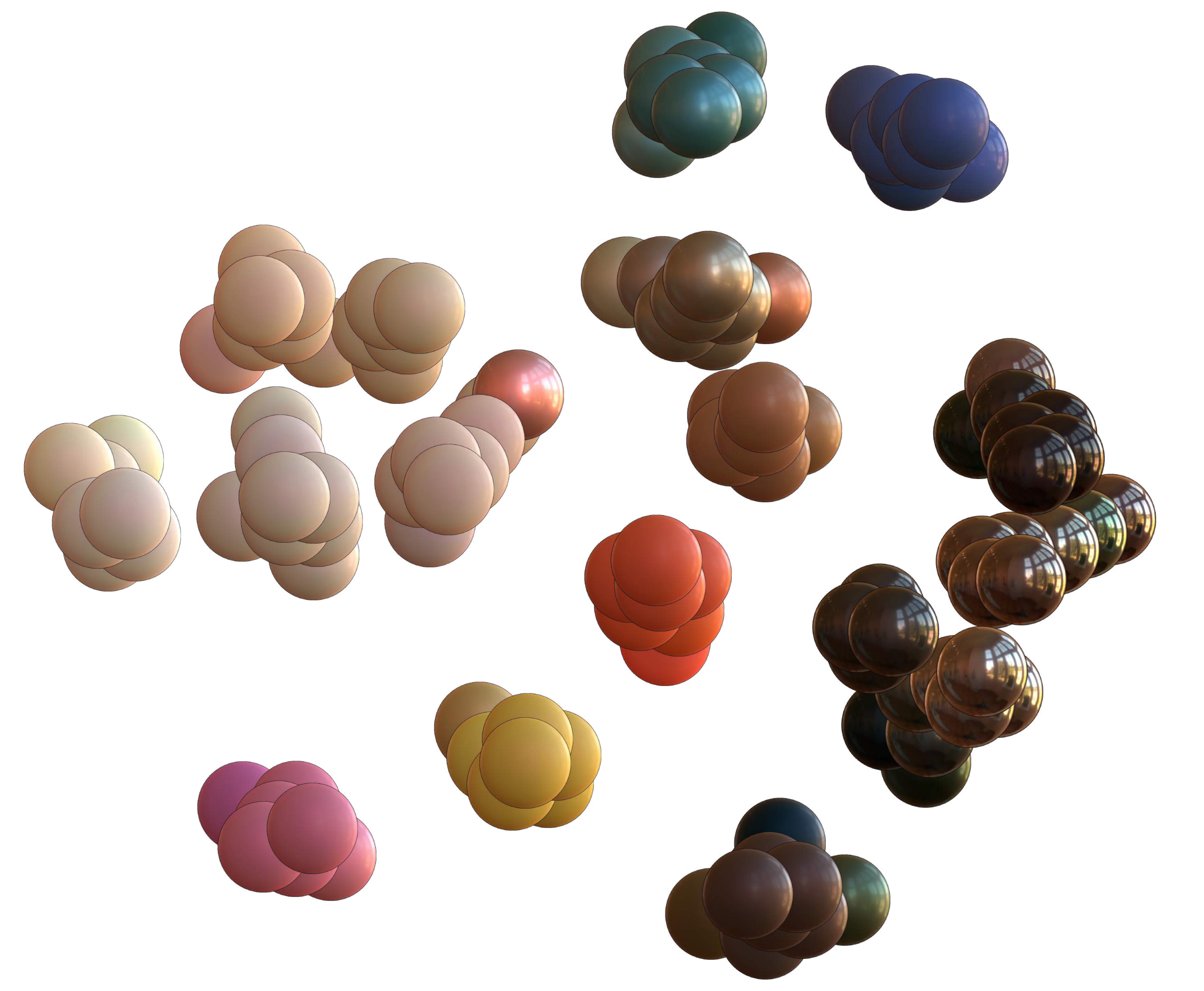}
   \caption{t-SNE clustering of the test embeddings with different sample sizes, including $N=8, 40, 160, 4\,000, 40\,000, 640\,000$.}
   \label{fig:tsne-vis-imputation}
\end{figure}

Considering our architecture, we understand that the gap between reconstruction results with sparse samples relies on the embeddings $z$ (latent vectors). We hence analyse the embeddings for the materials reconstructed with varying number of samples. In Figure \ref{fig:tsne-vis-imputation}, we visualize 
the t-SNE clustering of the test embeddings with different number of samples. It is visible that the embeddings of the same material reconstructed with different sample sizes lie close in the t-SNE space.

\subsubsection{Qualitative Comparison}\label{sec:qual_comp}
\begin{figure*}[t]
  \centering
\adjustbox{trim={0.\width} {.\height} {0.89\width} {.\height},clip}%
  {\includegraphics[width=0.9\linewidth]{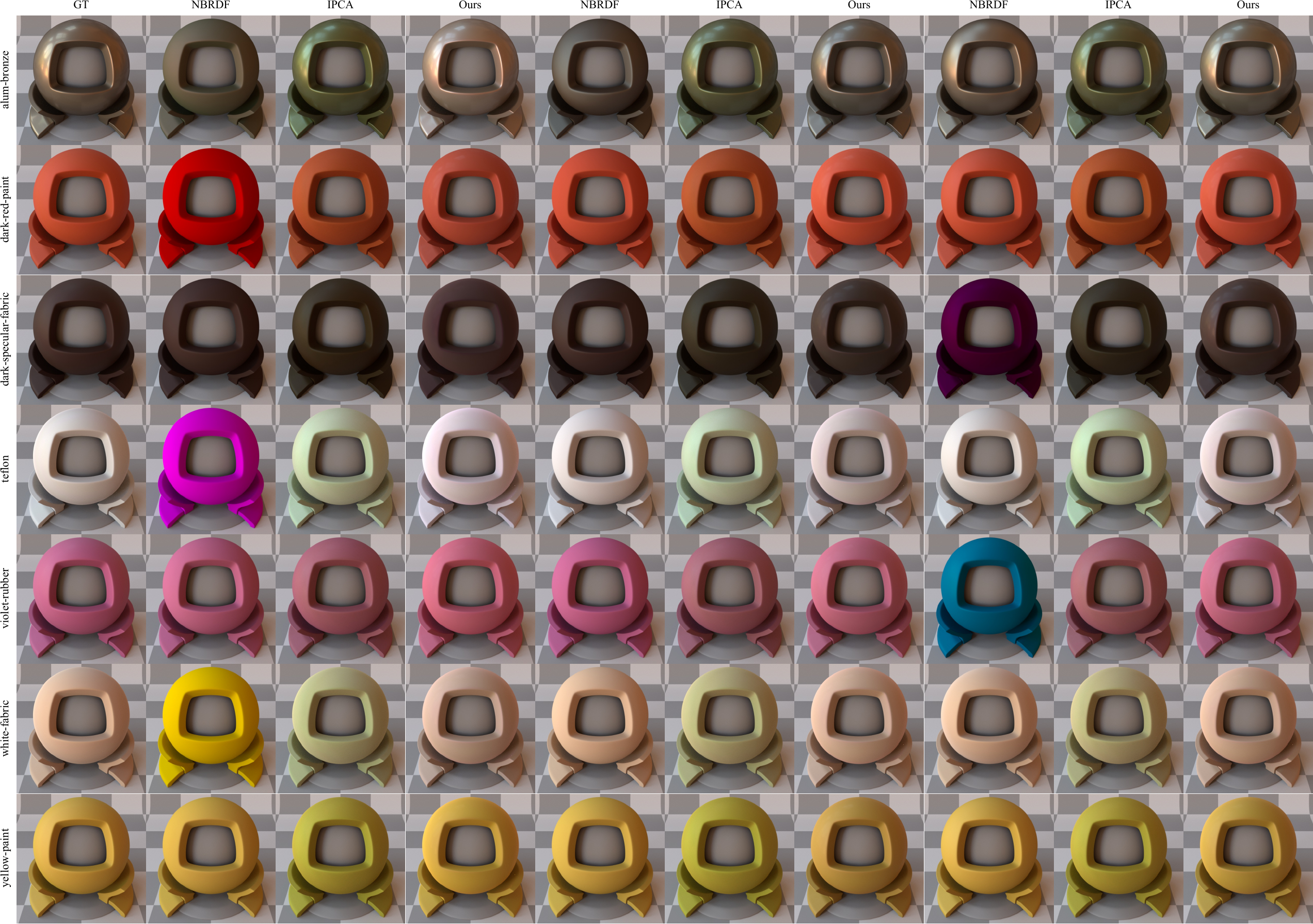}}
\adjustbox{trim={0.113\width} {.\height} {0.596\width} {.\height},clip}%
  {\includegraphics[width=0.9\linewidth]{fig/imp_comp_upt_3vals.pdf}}
\adjustbox{trim={0.41\width} {.\height} {0.298\width} {.\height},clip}%
  {\includegraphics[width=0.9\linewidth]{fig/imp_comp_upt_3vals.pdf}}
\adjustbox{trim={0.707\width} {.\height} {0.\width} {.\height},clip}%
  {\includegraphics[width=0.9\linewidth]{fig/imp_comp_upt_3vals.pdf}}

\adjustbox{trim={0.\width} {.\height} {0.89\width} {.06\height},clip}%
  {\includegraphics[width=0.9\linewidth]{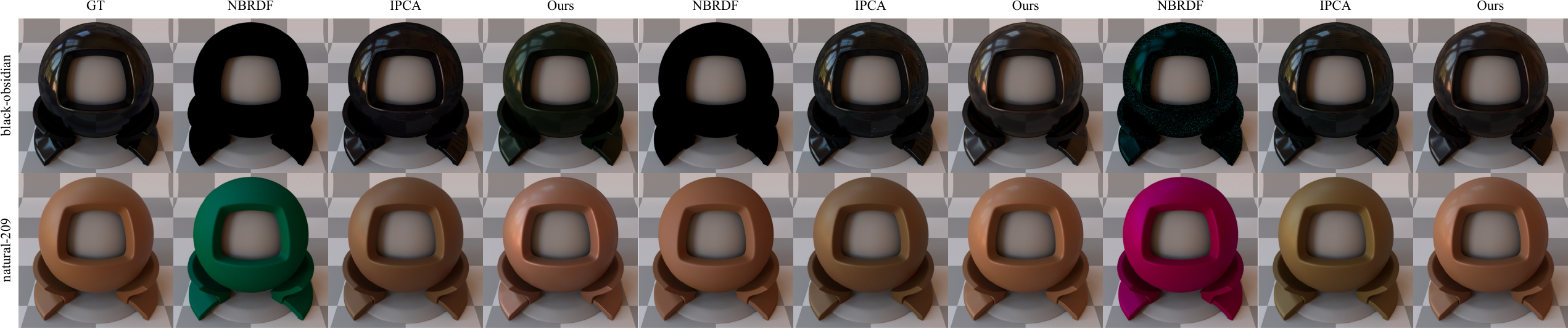}}
\adjustbox{trim={0.113\width} {.\height} {0.596\width} {.06\height},clip}%
  {\includegraphics[width=0.9\linewidth]{fig/imp_comp_upt_3vals_bad.pdf}}
\adjustbox{trim={0.41\width} {.\height} {0.298\width} {.06\height},clip}%
  {\includegraphics[width=0.9\linewidth]{fig/imp_comp_upt_3vals_bad.pdf}}
\adjustbox{trim={0.707\width} {.\height} {0.\width} {.06\height},clip}%
  {\includegraphics[width=0.9\linewidth]{fig/imp_comp_upt_3vals_bad.pdf}}
\\
\hspace{1.9cm}{\includegraphics[width=0.8\linewidth]{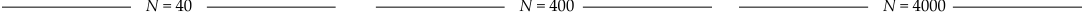}}
   \caption{Qualitative comparison results for reconstruction with small sample sizes. Thanks to the prior that our hypernetwork model learns for material appearance through training, it can accurately estimate the BRDFs of unseen materials and preserve the colors better than the baselines.}  


   \label{fig:imp_comp_upt}
\end{figure*}

\begin{figure*}[t]
  \centering
    {\includegraphics[width=0.35\linewidth]{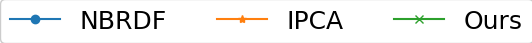}}\\
  {\includegraphics[width=0.32\linewidth]{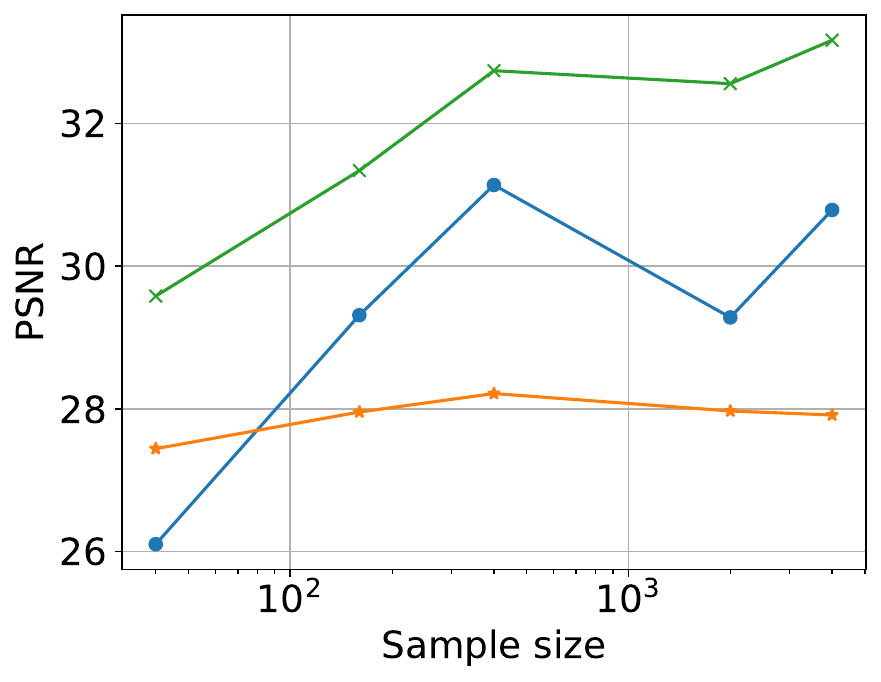}}
  {\includegraphics[width=0.32\linewidth]{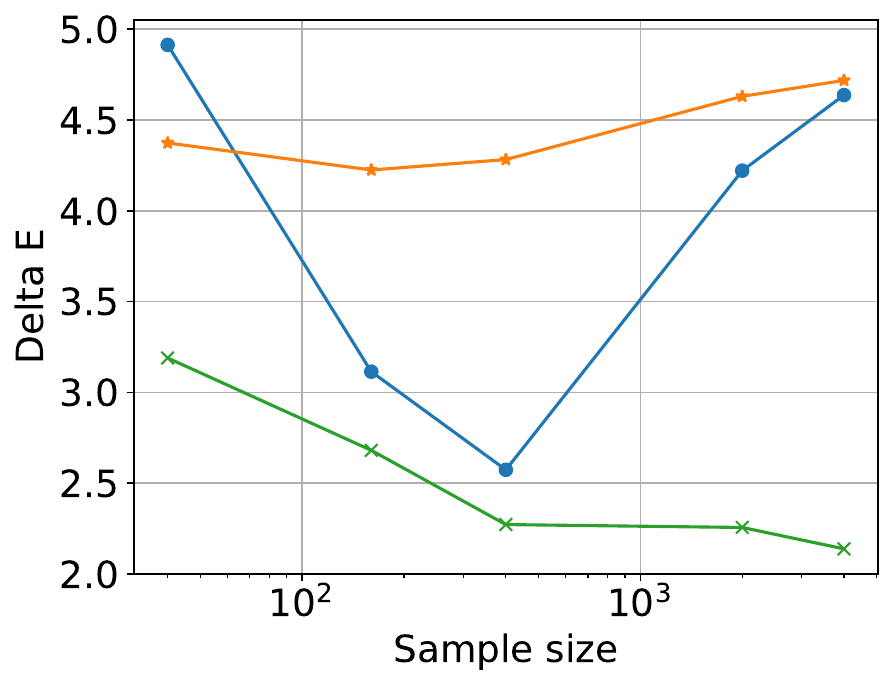}}
  \adjustbox{trim={0.\width} {.0\height} {0.\width} {.\height},clip}%
    {\includegraphics[width=0.32\linewidth]{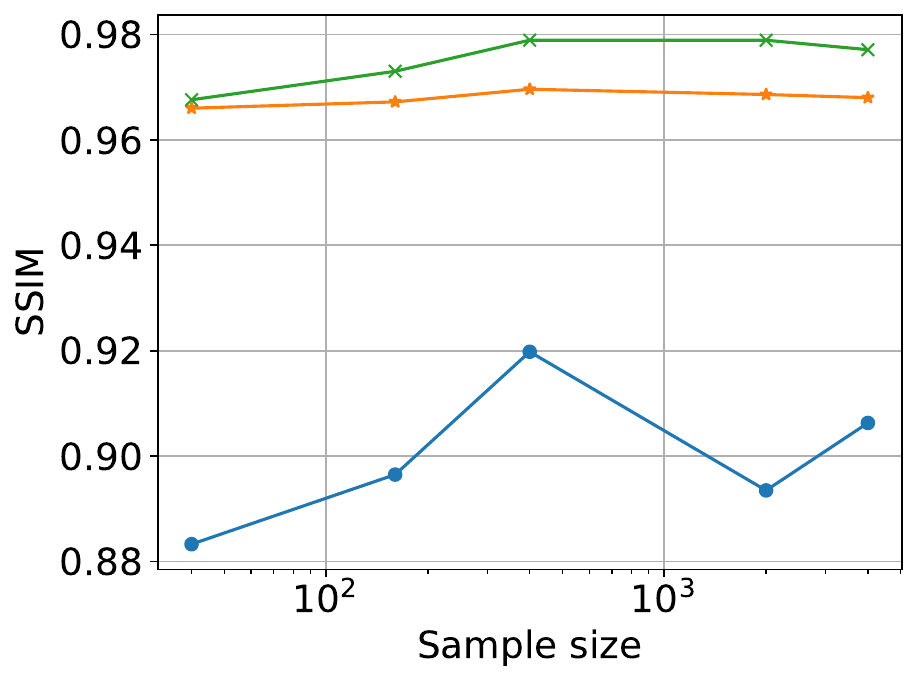}}
   \caption{Average PSNR, Delta E (CIE 2000), and SSIM results across different sample sizes. }
   \label{fig:imp_plots}
\end{figure*}

We first compare our results with the ground truth renderings for varying sample sizes. Figure \ref{fig:imp_comp_upt} shows that the hypernetwork can still reconstruct test materials even with $N = 40$ samples. However, as we reduce the sample number to 10, the reconstruction quality highly decreases. For additional results with different query sample sizes, please refer to our supplementary.



During inference time, hypernetwork fits the samples to a BRDF. Even with a few samples, it can still accurately reconstruct test materials. Its success comes from the fact that it learns a prior for material appearance through training with a dataset of multiple materials. Therefore, we validate our generalizable approach by comparing the hypernetwork with NBRDF \cite{sztrajman2021neural} and PCA with Log Relative Mapping (IPCA) \cite{nielsen2015optimal}.


NBRDF \cite{sztrajman2021neural} is designed to implicitly represent an individual BRDF. It can reconstruct materials with very high accuracy (Structural Similarity Index of around 0.995) when the sample size is high. It first trains the network on a material and then estimates the BRDF values of the same material. For a fair comparison, we first fit an NBRDF model to a material BRDF with the query sample size, then estimate the function with the same sample size. 

We also compare our results with a PCA-based strategy (IPCA). To represent the BRDF data, PCA-based methods  \cite{matusik2003data, ngan2006image} construct a matrix ${A} \in \mathbb{R}^{m \times n}$, where n = 180 × 90 × 90 × 3 = $4\,374\,000$ is the feature number, m = 80 is sample number. The matrix is later decomposed into its principal components via Singular Value Decomposition. PCA itself struggles with the decomposition of high dynamic range data. Therefore, \cite{nielsen2015optimal} applies the Log Relative Mapping (IPCA), which is the same as our pre-processing step (Section \ref{sec:pre-proc}). This improves the reconstruction qulity, offering competitive reconstructions against our method as shown in Figure \ref{fig:imp_comp_upt}.

For all methods including ours, samples are drawn from a uniform distribution over Rusinkiewicz coordinates. For IPCA results, we split train and test dataset in the same way as ours, i.e., 80 MERL materials for train and 20 for test, and use all available samples for learning the principal components. To reconstruct the test materials from sparse samples, we first run a least-squared-error optimization and predict the weights of the material for the corresponding principal components. To decide the number of components, we ran an additional ablation study, where 
we computed the average mean squared error over the test dataset for the reconstruction from sparse samples ($N = 40$). We observed that ($N_{PC} = 8$) gives the minimum error on the test dataset. Hence, we choose the number of principal components as $N_{PC} = 8$ and keep it same for sparse sampling results.

We keep the range for sparse sample numbers between $N = 40$ and $N = 4000$. Figure \ref{fig:imp_comp_upt} shows the rendering results for the minimum and maximum number of our range. Compared to both IPCA and NBRDF, our hypernetwork can capture the appearances more precisely. Although our approach has difficulty estimating specular components (last two rows in Figure~\ref{fig:imp_comp_upt}), overall it offers reconstructions with much higher accuracy.

We observe that the diffuse colors of some reconstructed materials by NBRDF can be completely off (natural-209, teflon) due to function fitting with sparse samples. Hypernetwork, on the other hand, can preserve diffuse colors better thanks to the prior it learns through training.

\subsubsection{Quantitative Evaluation}

We compare our method quantitatively with the aforementioned techniques in multiple image-based error metrics through rendering results. The metrics include peak signal-to-noise ratio (PSNR), Delta E (CIE 2000), and structural similarity index (SSIM). We take the average over the test dataset for each metric. We plot the metric results across five different sample sizes (40, 160, 400, 2000, 4000). Since we optimize IPCA on the test dataset for the $N = 40$ case, it offers more competitive results. Nevertheless, our method can reconstruct the BRDFs of unseen materials more precisely. Figure \ref{fig:imp_plots} shows that across all sample sizes, the hypernetwork attains superior performance in terms of PSNR, Delta E (CIE 2000) and SSIM. 

Additionally, Table \ref{table: ours_diff_samples} shows that expanding the training set, even with materials captured from a different point-based setup, helps improve the performance of the hypernetwork. Also, note that compared to MERL materials, the colors of RGL materials are more saturated, which could explain a slight increase in Delta E.

 \begin{table*}
    \centering
    \caption{Hypernetwork sparse reconstruction - Average metric results across varying sample sizes ($N$) over the test set. We highlight \colorbox{blue!25}{best} and \colorbox{orange!25}{second best} results.}
    
    {\begin{tabular}{l@{\hskip 0.4in}c@{\hskip 0.2in}c@{\hskip 0.2in}c@{\hskip 0.1in}|@{\hskip 0.1in}c@{\hskip 0.2in}c@{\hskip 0.2in}c}\toprule
    
    
& \multicolumn{3}{c}{MERL} & \multicolumn{3}{c}{MERL + RGL}
\\\cmidrule(lr){2-4}\cmidrule(lr){5-7}
  $N$ & PSNR\textuparrow & Delta E\textdownarrow & SSIM\textuparrow & PSNR\textuparrow & Delta E\textdownarrow & SSIM\textuparrow \\

 \toprule

$40$ & 29.581 & 3.189 & 0.968 & 30.018 & 3.112 & 0.963\\
$160$ & 31.341 & 2.681 & 0.973 & 31.929 & 2.454 & 0.962\\
$400$ & 32.743 & 2.272 & \cellcolor{blue!25} 0.979 & 33.855 & 2.432 & 0.977\\
$2000$ & \cellcolor{blue!25} 34.527 & \cellcolor{orange!25}2.256 & \cellcolor{blue!25} 0.979 & \cellcolor{blue!25} 34.527 & \cellcolor{orange!25} 2.243 & \cellcolor{blue!25} 0.983\\
$4000$ & \cellcolor{orange!25} 33.170 &  \cellcolor{blue!25} 2.138 & \cellcolor{orange!25} 0.977 & \cellcolor{orange!25} 34.355 & \cellcolor{blue!25} 2.166 & \cellcolor{orange!25} 0.982\\

\bottomrule
    \end{tabular}\par}
    \label{table: ours_diff_samples}
\end{table*}

\subsection{Compression}\label{sec:compression}
The high capacity of the hypernetwork also allows the compression of the densely sampled BRDF data into low-dimensional latent representations. The hypernetwork can process highly compact BRDF embeddings, and once decoded, reconstructs the BRDF data precisely. We compare our method with Neural Processes (NPs) \cite{zheng2021compact}, the state-of-the-art BRDF compression method, and show that our hypernetwork model overall performs better in all three metrics. 


To compare our method with NPs, we overfit our model to the mixed dataset of MERL and isotropic RGL materials, consisting of 151 materials in total. Since the latent dimension of NPs is 7D, we also train our network with 7D latent space. It is worth mentioning that NPs cause 
invalid sample values in certain MERL materials, blacking some parts of the renderings. In contrast, our method consistently decompresses materials with high reconstruction accuracy; see Table \ref{table: oursvsnps}. Rendering results can be found in our supplementary (Compression.pdf).

\begin{table}
    \centering
    \caption{Compression - Average metric results over the renderings of the entire MERL dataset. We highlight \colorbox{blue!25}{best} and \colorbox{orange!25}{second best} results.}

    {%
    {\begin{tabular}{l@{\hskip 0.5in}c@{\hskip 0.3in}c@{\hskip 0.3in}c}\toprule


  &  PSNR \textuparrow & Delta E \textdownarrow & SSIM \textuparrow \\
 \toprule
 Ours (40D) & \cellcolor{blue!25} 47.682 & \cellcolor{blue!25} 0.567 & \cellcolor{blue!25} 0.994\\
 Ours (7D) & \cellcolor{orange!25} 47.492 & \cellcolor{orange!25} 0.574 & \cellcolor{blue!25} 0.994\\
 NPs & 46.125 & 2.424 & 0.935\\
 IPCA & 29.892 & 3.315 & 0.979\\

\bottomrule
    \end{tabular}\par}}
    \label{table: oursvsnps}
\end{table}


\begin{figure*}[h]
  \centering
  {\includegraphics[width=0.9\linewidth]{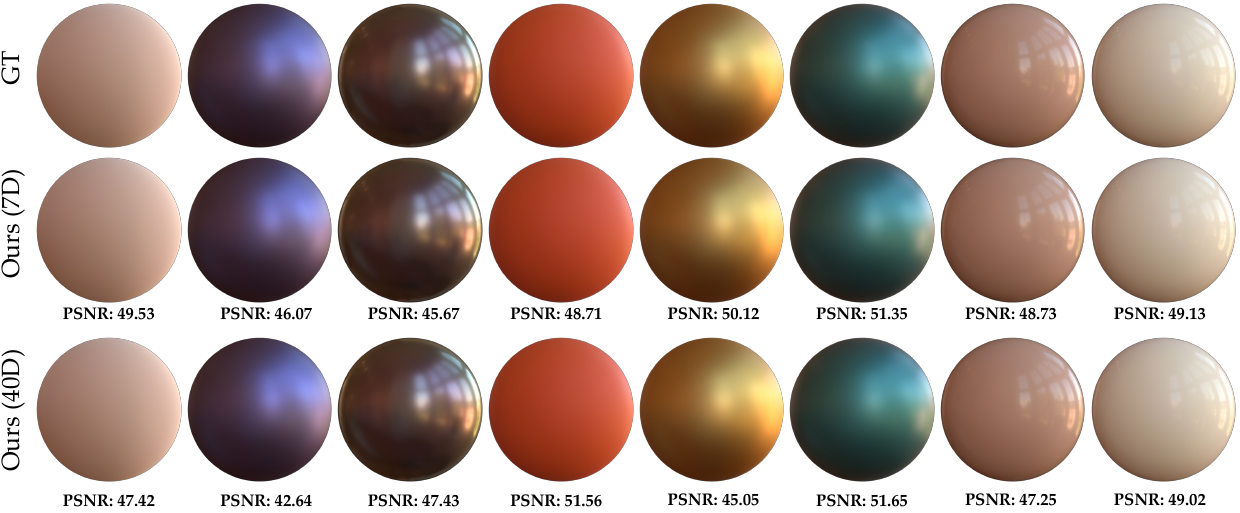}}
   \caption{Reconstruction results for BDRF compression (GT: ground truths).}
   \label{fig:comp-fig}
\end{figure*}

\subsection{BRDF Editing}

\begin{figure*}[h]
  \centering
   \includegraphics[width=0.9\linewidth]{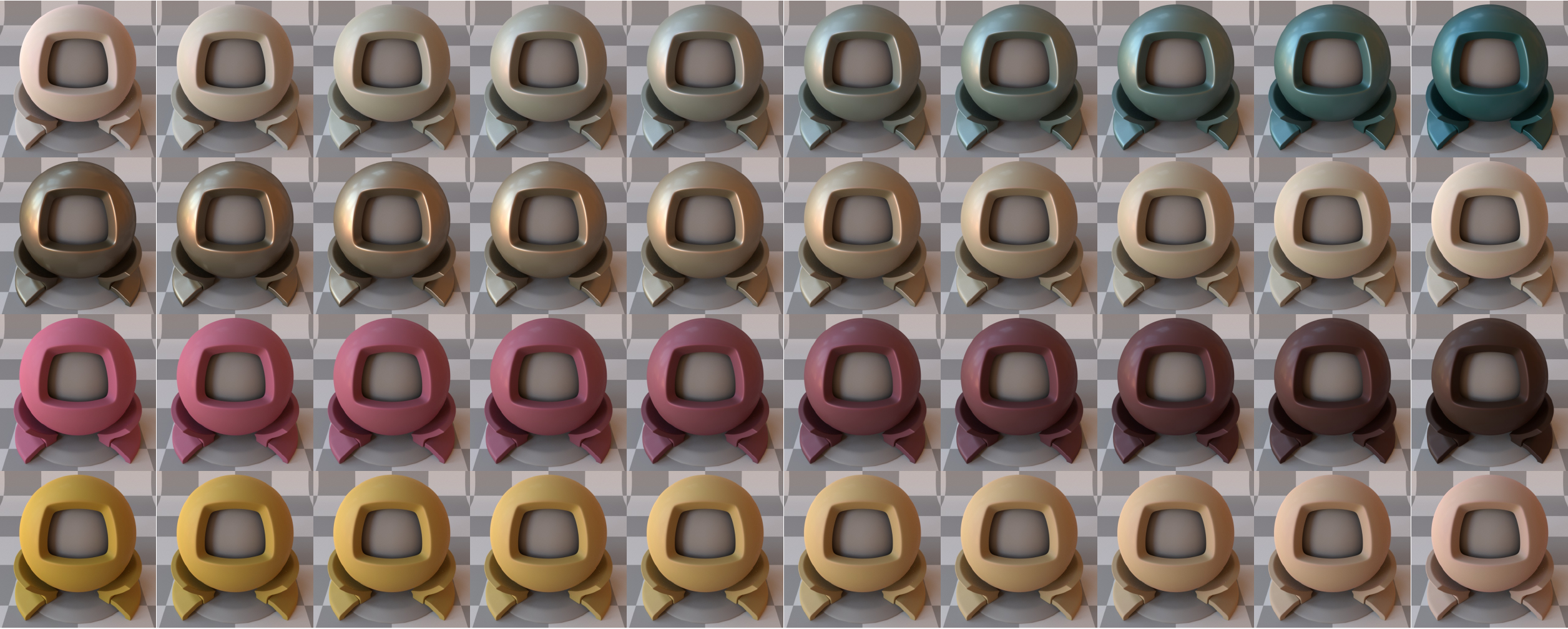}

   \caption{BRDF editing through linear interpolation between the embeddings of two materials.}
   \label{fig:interpolation}
\end{figure*}

Compared to analytic BRDFs that have a fixed number of parameters to tweak, editing measured BRDFs is rather a nontrivial task due to irregular data structure. On the other hand, our network is capable of BRDF editing thanks to its representation of materials in a low-dimensional space. We can easily reconstruct various appearances by linearly interpolating between the embeddings of two different materials. Figure \ref{fig:interpolation} shows newly-reconstructed materials through linear interpolation between two different embeddings from the reconstructed MERL materials. 


\subsection{Limitations and Future Work}\label{sec:limits}
\paragraph{Specular components:} Our network struggles with the estimation of specular components as shown in Figure~\ref{fig:imp_comp_upt} (last two rows). It is likely because of the high gap between the values of diffuse that are close to zero and the values of specular components that are arbitrarily high. We consider a separate estimation pipeline for each component within the network to improve the results.



\paragraph{BRDF editing:} Our BRDF editing approach is rather non-intuitive with an interpolation approach. Finding a map between the embeddings and certain attributes of the materials, such as diffuse/specular colors or haziness, can lead to more interactive BRDF editing. We plan to map our latent space to material parameter space so that users can easily edit materials through more interpretable parameters.

\paragraph{SVBRDF representations:} In this work, we focused on the task of generalizable neural representations for spatially uniform BRDFs. As the future work, we plan to extend our model to SVBRDF representations.





%% file: sec/5_conclusion.tex
\section{Conclusion}\label{sec:conc}

We presented a hypernetwork model for neural BRDF representation, which is suitable for point-based BRDF acquisition setups. Representing the discrete BRDF values of a material as a continuous generalizable neural field, our method offers (1) accurate reconstructions of unseen materials from a sparse set of samples and (2) compression of the highly densed BRDF values into compact latent representations. 

Thanks to the set encoder that enables an arbitrary number of input samples and the neural field, hyponet, that offers the nonlinear built-in interpolation, we showed that our hypernetwork model can capture the BRDFs of test materials better than our baselines. We also illustrated our model's compression capacity on the entire MERL dataset, showing consistent results across varying materials.